\documentclass[twocolumn,floatfix,preprintnumbers,superscriptaddress]{revtex4}

\usepackage[utf8]{inputenc}
\usepackage[colorlinks=true,citecolor=blue,linkcolor=blue]{hyperref}
\usepackage[normalem]{ulem}
\usepackage{url}
\usepackage{graphicx,wrapfig,float,slashed,cancel}
\usepackage{amsmath,amssymb,epsfig,graphicx,xcolor,stmaryrd}
\usepackage{bm}
\usepackage{enumitem}
\usepackage{xcolor}

\definecolor{darkblue}{RGB}{1, 90, 173}

\begin{document}


\title{Heavy-light hybrid mesons with different spin-parities}

\author{B. Barsbay\textsuperscript}
\email{bbarsbay@dogus.edu.tr}
\affiliation{Department of Physics, Do\v{g}u\c{s} University, Dudullu-\"{U}mraniye, 34775
Istanbul, Turkey}

\author{K. Azizi} %
\email{kazem.azizi@ut.ac.ir}
\thanks{Corresponding Author}
\affiliation{Department of Physics, University of Tehran, North Karegar Ave. Tehran 14395-547, Iran}
\affiliation{Department of Physics, Do\v{g}u\c{s} University, Dudullu-\"{U}mraniye, 34775
Istanbul, Turkey}
\affiliation{School of Particles and Accelerators, Institute for Research in Fundamental
Sciences (IPM) P.O. Box 19395-5531, Tehran, Iran}

\author{H.~Sundu}
\email{ hayriye.sundu@kocaeli.edu.tr}
\affiliation{Department of Physics, Kocaeli University, 41380 Izmit, Turkey}

\date{\today}

\preprint{}

\begin{abstract}
	The spectroscopic parameters of the heavy-light hybrid mesons with different spin-parities and different quark contents are investigated in the framework of the QCD sum rule method. The mass and current coupling of these states are calculated  by taking into account the quark, gluon and mixed vacuum condensates up to dimension 10. The obtained results are compared with the existing QCD Laplace sum rule predictions. The results of mass and current coupling  for all the considered channels  are obtained to be stable and reliable. Our results can be useful for future experimental searches  as well as  theoretical studies of different parameters related to the hybrid mesons and their decays and interactions with other particles.
\end{abstract}


\maketitle

\section{Introduction} \label{sec:intro}

The standard hadrons, i.e., mesons and baryons are respectively made of two and three valence quarks, which  strongly interact through gluon exchange. The field theory of these interactions is the Quantum Chromodynamics (QCD) and they are categorized using the quark model. Both the QCD and quark model do not exclude existing of other states with different configurations. Hence, it was already suggested that, in addition to the conventional hadrons, there may exist particles composed of various combinations of quarks and gluons. Some of such states were observed in numerous experiments. These states are particles including four-quark XYZ tetraquarks, hadronic molecules composed of mesons and baryons, pentaquarks, glueballs, hybrids, etc., which are nominated as exotic states. Exploring the existence and properties of such exotic states is one of the most interesting research topics of high energy physics. In the past two decades, with the development of experimental facilities, starting with the observation of the $X(3872)$ ~\cite{Choi:2003ue}, we witness the observations of many exotic hadrons~ (for more information see for instance Refs. \cite{D0:2016mwd,LHCb:2016axx,LHCb:2016nsl,Belle:2007umv,Belle:2014wyt,LHCb:2015yax,LHCb:2016lve,LHCb:2019kea,Chen:2022asf,Meng:2022ozq}).

An important category in the exotic hadrons is hybrid hadrons containing valence gluon(s), besides valence quarks. The existence of hybrids was first predicted  by Jaffe and Johnson in 1976~\cite{Jaffe:1975fd}. Understanding the inner organizations and properties of these hadrons may support their future investigations. Besides, they may carry new information on QCD. Therefore, the hybrids hadrons have been studied in the framework of different theoretical methods such as the constituent gluon model~\cite{Horn:1977rq}, the flux-tube model~\cite{Barnes:1995hc}, lattice QCD~\cite{Perantonis:1990dy,Liu:2005rc,HadronSpectrum:2012gic,Liu:2011rn,Cheung:2016bym} and QCD sum rules~\cite{Govaerts:1984hc, Govaerts:1985fx, Govaerts:1986pp, Zhu:1998ki, Narison:2009vj, Qiao:2010zh, Harnett:2012gs, Berg:2012gd, Chen:2013pya, Kleiv:2014kua,Palameta:2017ols, Palameta:2018yce}. Unfortunately, the predictions for the masses of the hybrids acquired within these approaches are inconsistent with each other. Moreover, most of the studies have been carried out on heavy quarkonium hybrids and there are less studies devoted to the investigations of the  properties of heavy-light hybrids, which make an important category of these states. The first investigations of heavy-light hybrids using QCD sum rule were performed by Govaerts, Reinders, and Weyers in 1985~\cite{Govaerts:1985fx}. Thus, more studies are needed to clarify the physical properties of this category of heavy-light hybrids, which may be in agenda of future experiments.

Motivated by this situation, in this article, we are going to calculate the mass and current coupling of the scalar, pseudoscalar, vector and axial-vector heavy-light hybrid mesons with all possible quantum numbers and quark contents in the framework of the Borel QCD sum rule method. This method is one of the powerful and predictive non-perturbative approaches in hadron physics~\cite{Shifman:1978bx, Shifman:1978by, Reinders:1984sr, Narison:1989aq}.  This approach has been successfully applied to not only the standard hadrons but also to the exotics  (see for instance Refs.  \cite{Wang:2018ntv,Wu:2018xdi,Voloshin:2018vym,Cao:2018vmv, Agaev:2021vur,Agaev:2020zad,Agaev:2020mqq,Sundu:2018uyi,Agaev:2019qqn,Wang:2021itn,Azizi:2021utt,Wang:2020rdh,Wang:2018waa,Wang:2019iaa,Wang:2019hyc,Azizi:2021pbh}) and the obtained results well explain the existing experimental data. 

The spectroscopic parameters of the  heavy-light hybrid mesons with various quark contents have already been calculated in Ref. \cite{Ho:2016owu} using  QCD Laplace sum rules. As a result of analyses, the states with $J^{P(C)}\in \{0^{+(+)},1^{-(-)},1^{+(-)}\}$  were led to stable results on the mass and current coupling constant, while for the states with  the $J^{P(C)}\in \{0^{+(-)},1^{+(+)},1^{-(+)}\}$ quantum numbers the analyses were led to unstable predictions. In the present study we construct various currents,  interpolating the heavy-light hybrid mesons of different possible spin-parities and quark contents, to investigate the spectroscopic properties of different kinds of the heavy-light hybrid mesons. The scalar and vector states as well as the pseudoscalar and axial-vector states couple to the same currents. We separate contribution of these states and calculate the mass and current coupling of all  these hybrid mesons. We get stable and reliable results in all channels considered in the present study.

This work is organized in the following manner. In sec. \ref{sec:Mass}, we  derive QCD sum rules for the mass and current coupling of the heavy-light hybrid mesons of different spin-parities. In sec. \ref{sec:Numeric}, we  numerically analyze the obtained sum rules  and present our predictions for the mass and current coupling of the considered states. Section \ref{sec:Conc} is reserved for summary and conclusion.

\section{Mass and current coupling of the heavy-light hybrid mesons}
\label{sec:Mass}
To calculate the mass and current coupling of the heavy-light hybrid mesons, we start with the two-point correlation function
\begin{align}
  \Pi _{\mu \nu }(q)=&i\int d^{4}xe^{iqx}\langle 0|\mathcal{T}\{J_{\mu
    }(x)J_{\nu }^{\dagger }(0)\}|0\rangle \nonumber \\
  =&\frac{q_{\mu}q_{\nu}}{q^2}\Pi_{S(PS)}(q^2) 
   + \left(\frac{q_{\mu}q_{\nu}}{q^2}-g_{\mu\nu}\right)\Pi_{V(AV)}(q^2),  \label{eq:CorrF1}
\end{align}

where $J_{\mu }(x)$ is the interpolating current for a vector and an axial-vector hybrid states. The currents are taken as
\begin{eqnarray}
&&J_{\mu }^{1}(x)=\frac{g_s}{2} \overline{Q}_{a}(x)\gamma
_{\theta}\gamma _{5} \frac{\lambda _{ab}^{n}}{2} \tilde{G}_{\mu\theta}^{n}(x)q_{b} (x),  \label{eq:Curr1}
\end{eqnarray}%
for $0^{+(-)}$ and $1^{-(-)}$ quantum numbers,
\begin{eqnarray}
&&J_{\mu }^{2}(x)=g_s \overline{Q}_{a}(x)\gamma
_{\theta}\gamma _{5} \frac{\lambda _{ab}^{n}}{2} G_{\mu\theta}^{n}(x)q_{b} (x),  \label{eq:Curr2}
\end{eqnarray}%
for $0^{-(-)}$ and $1^{+(-)}$ quantum numbers,
\begin{eqnarray}
&&J_{\mu }^{3}(x)=\frac{g_s}{2} \overline{Q}_{a}(x)\gamma
_{\theta} \frac{\lambda _{ab}^{n}}{2} \tilde{G}_{\mu\theta}^{n}(x)q_{b} (x),  \label{eq:Curr3}
\end{eqnarray}%
for $0^{-(+)}$ and $1^{+(+)}$ quantum numbers, and
\begin{eqnarray}
&&J_{\mu }^{4}(x)=g_s \overline{Q}_{a}(x)\gamma
_{\theta} \frac{\lambda _{ab}^{n}}{2} G_{\mu\theta}^{n}(x)q_{b} (x), \label{eq:Curr4}
\end{eqnarray}%
for the states with $0^{+(+)}$ and $1^{-(+)}$ quantum numbers.
In the above currents, the $g_{s}$ is the strong coupling constant, $a,b= 1,2,3$, $\lambda^n$ with $n= 1,2, \cdots, 8$ are the Gell-Mann matrices; $\tilde{G}_{\mu\theta}^{n}(x) = \epsilon_{\mu\theta\alpha\beta} G_{\alpha \beta}^{n}(x)/ 2$ is the dual field strength of $G_{\mu\theta}^{n}(x)$; and $q=u, d, s$ and $Q=c, b$ are light and heavy quarks, respectively.

In this work, we will reference each of the $\Pi _{\mathrm{S(PS)}}(q^{2})$ and $\Pi _{\mathrm{V(AV)}}(q^{2})$ according to the chosen quantum numbers of the heavy-light hybrid meson. In order to derive desired sum rules, we write down the correlation function $\Pi
_{\mu \nu }(q)$ using the mass and current coupling of the heavy-light hybrid meson. Saturating the correlation function with  complete sets of states with quantum numbers of hybrid state and performing the integral over $x$ in Eq.\ (\ref{eq:CorrF1}), we get the following result:
\begin{eqnarray}
&&\Pi _{\mu \nu }^{\mathrm{Phys}}(q)=\frac{\langle 0|J_{\mu }|H_{S(PS)}(q)\rangle
\langle H_{S(PS)}(q)|J_{\nu }^{\dagger }|0\rangle }{m_{H_{S(PS)}}^{2}-q^{2}}   \notag \\
&&+\frac{\langle 0|J_{\mu }|H_{V(AV)}(q)\rangle
\langle H_{V(AV)}(q)|J_{\nu }^{\dagger }|0\rangle }{m_{H_{V(AV)}}^{2}-q^{2}}+\ldots,  \notag \\
&&  \label{eq:PhysSide}
\end{eqnarray}%
where $m_{H_{S(PS)}}$ and $m_{H_{V(AV)}}$ are the masses of the scalar (pseudoscalar) and vector (axial-vector) hybrid states. As is seen,  the states $ S $ and $ V $ mix with each other and  $ PS $ and $ AV $ come together in the physical representation. 
The function can be simplified by introducing the matrix elements
\begin{equation}
\langle 0|J_{\mu }|H_{S(PS)}(q)\rangle =q_{\mu}f_{H_{S(PS)}}\,
\label{eq:Mel1}
\end{equation}%
and 
\begin{equation}
\langle 0|J_{\mu }|H_{V(AV)}(q)\rangle =m_{H_{V(AV)}}f_{H_{V(AV)}}\varepsilon _{\mu },
\label{eq:Mel1}
\end{equation}%
where $\varepsilon _{\mu }$ is the polarization vector of the vector (axial-vector) hybrid state. $f_{H_{S(PS)}}$ and $f_{H_{V(AV)}}$ are the current couplings of the scalar (pseudoscalar) and vector (axial-vector) mesons. In terms of the parameters entered, the function $\Pi _{\mu \nu }^{\mathrm{Phys}}(q)$ takes the form
\begin{align}
\Pi _{\mu \nu }^{\mathrm{Phys}}(q)=&\frac{f_{H_{S(PS)}}^{2}}{m_{H_{S(PS)}}^{2}-q^{2}}q_{\mu }q_{\nu }+\frac{m_{H_{V(AV)}}^{2}f_{H_{V(AV)}}^{2}}{m_{H_{V(AV)}}^{2}-q^{2}} \nonumber \\
&\times\left( -g_{\mu \nu }+\frac{q_{\mu }q_{\nu }}{q^{2}}\right) +\cdots. 
\label{eq:Phys}
\end{align}%
As we previously mentioned, the correlator $\Pi _{\mu \nu }^{\mathrm{Phys}}(q)$ includes $\mathrm{S}$ and $\mathrm{V}$ contributions as well as $\mathrm{PS}$ and $\mathrm{AV}$ contributions, simultaneously. To separate the $\mathrm{S}$ ($\mathrm{PS}$) contribution, Eq.\ (\ref{eq:Phys}) is multiplied by $q_{\mu }q_{\nu }/{q^{2}}$ which leads to
\begin{equation}
\frac{q_{\mu }q_{\nu }}{q^{2}}\Pi _{\mu \nu }^{\mathrm{Phys}}(q)=-f_{H_{S(PS)}}^{2}+\frac{m_{H_{S(PS)}}^{2}f_{H_{S(PS)}}^{2}}{m_{H_{S(PS)}}^{2}-q^{2}} +\cdots.
\label{eq:Phys1}
\end{equation}%
To find the contributions of the vector and axial-vector states, it is enough to chose the structure $g_{\mu \nu }$ which contains only the contribution of $\mathrm{V(AV)}$ particle and it is free of the contamination by $\mathrm{S(PS)}$ state:
\begin{equation}
\Pi _{\mu \nu }^{\prime\mathrm{Phys}}(q)=\frac{m_{H_{V(AV)}}^{2}f_{H_{V(AV)}}^{2}}{m_{H_{V(AV)}}^{2}-q^{2}}\times\left( -g_{\mu \nu }\right) +\cdots. 
\label{eq:Physp}
\end{equation}%
The Borel transformation with respect to $q^{2}$ applied to $\Pi _{\mu \nu }^{\prime\mathrm{Phys}}(q)$ and $q_{\mu }q_{\nu }/{q^{2}}\Pi _{\mu \nu }^{\mathrm{Phys}}(q)$ leads to the final forms of the physical sides:
\begin{equation}
\mathcal{B}_{q^{2}}\Pi _{\mu \nu }^{\prime\mathrm{Phys}%
}(q)=m_{H_{V(AV)}}^{2}f_{H_{V(AV)}}^{2}e^{-m_{H_{V(AV)}}^{2}/M^{2}}\left( -g_{\mu \nu }\right) +\cdots,  \label{eq:CorBor1}
\end{equation}
and
\begin{equation}
\mathcal{B}_{q^{2}}\frac{q_{\mu }q_{\nu }}{q^{2}}\Pi _{\mu \nu }^{\mathrm{Phys}}(q)=m_{H_{S(PS)}}^{2}f_{H_{S(PS)}}^{2}e^{-m_{H_{S(PS)}}^{2}/M^{2}}\ +\cdots .  \label{eq:CorBor2}
\end{equation}

The QCD side of the sum rules, $\Pi _{\mu \nu }^{\mathrm{OPE}}(q)$, has to be calculated in the operator product expansion (OPE) with certain accuracy. To get $\Pi _{\mu \nu }^{\mathrm{OPE}}(q)$, we calculate the correlation function using explicit forms of the currents $J_{\mu }^{1}(x)$, $J_{\mu }^{2}(x)$, $J_{\mu }^{3}(x)$ and $J_{\mu }^{4}(x)$. As a result, we express $\Pi _{\mu \nu }^{\mathrm{OPE}}(q)$ for each current in terms of the heavy and light quark propagators as well as the gluon propagator/condensate, 
\begin{eqnarray}
&&\Pi _{\mu \nu }^{\mathrm{OPE}_{(1,1)}}(q)=\frac{g_s^{2}\epsilon_{\mu\theta\alpha\beta}\epsilon_{\nu\delta\alpha^{\prime }\beta^{\prime }}}{4}i\int d^{4}xe^{iqx}  \notag \\
&&\times \langle 0|G^n_{\alpha \beta}(x) G^m_{\alpha^{\prime }\beta^{\prime }}(0)|0 \rangle \mathrm t^{n}[a,b] \mathrm t^{m}[a^{\prime },b^{\prime }] \notag \\
&&\times \mathrm{Tr}\left[S_{Q}^{a^{\prime }a}(-x)\gamma _{\theta }\gamma _{5}S_{q}^{bb^{\prime }}(x)\gamma _{\delta }\right],  \notag \\
&&  \label{eq:OPE1}
\end{eqnarray}%
\begin{eqnarray}
&&\Pi _{\mu \nu }^{\mathrm{OPE}_{(2,2)}}(q)=g_s^{2}i\int d^{4}xe^{iqx}\langle 0|G^n_{\mu \theta}(x) G^m_{\nu \delta}(0)|0 \rangle  \notag \\
&&\times \mathrm t^{n}[a,b] \mathrm t^{m}[a^{\prime },b^{\prime }] \mathrm{Tr}\left[S_{Q}^{a^{\prime }a}(-x)\gamma _{\theta }\gamma _{5}S_{q}^{bb^{\prime }}(x)\gamma _{5}\gamma _{\delta }\right],  \notag \\
&&  \label{eq:OPE2}
\end{eqnarray}%
\begin{eqnarray}
&&\Pi _{\mu \nu }^{\mathrm{OPE}_{(3,3)}}(q)=-\frac{g_s^{2}\epsilon_{\mu\theta\alpha\beta}\epsilon_{\nu\delta\alpha^{\prime }\beta^{\prime }}}{4}i\int d^{4}xe^{iqx}  \notag \\
&&\times \langle 0|G^n_{\alpha \beta}(x) G^m_{\alpha^{\prime }\beta^{\prime }}(0)|0 \rangle \mathrm t^{n}[a,b] \mathrm t^{m}[a^{\prime },b^{\prime }] \notag \\
&&\times \mathrm{Tr}\left[S_{Q}^{a^{\prime }a}(-x)\gamma _{\theta }S_{q}^{bb^{\prime }}(x)\gamma _{\delta }\right],  \notag \\
&&  \label{eq:OPE3}
\end{eqnarray}%
and
\begin{eqnarray}
&&\Pi _{\mu \nu }^{\mathrm{OPE}_{(4,4)}}(q)=-g_s^{2}i\int d^{4}xe^{iqx}\langle 0|G^n_{\mu \theta}(x) G^m_{\nu \delta}(0)|0 \rangle  \notag \\
&&\times \mathrm t^{n}[a,b] \mathrm t^{m}[a^{\prime },b^{\prime }] \mathrm{Tr}\left[S_{Q}^{a^{\prime }a}(-x)\gamma _{\theta }S_{q}^{bb^{\prime }}(x)\gamma _{\delta }\right],  \notag \\
&&  \label{eq:OPE4}
\end{eqnarray}%
where $t^n={\lambda^n}/2$.
In the above formulas, $S_{q}^{ab}(x)$ and $S_{Q}^{ab}(x)$ are propagators of $q(u,d,s)$ and $c,b$-quarks.
In the present work, for the light quark propagator, $S_{q}^{ab}(x)$, we
employ the following expression
\begin{eqnarray}
&&S_{q}^{ab}(x)=i\frac{\slashed x\delta _{ab}}{2\pi ^{2}x^{4}}-\frac{%
m_{q}\delta _{ab}}{4\pi ^{2}x^{2}}-\frac{\langle \overline{q}q\rangle }{12}%
\left( 1-i\frac{m_{q}}{4}\slashed x\right) \delta _{ab}  \notag \\
&&-\frac{x^{2}}{192}\langle \overline{q}g_{s}\sigma Gq\rangle \left( 1-i%
\frac{m_{q}}{6}\slashed x\right) \delta _{ab}-\frac{\slashed xx^{2}g_{s}^{2}%
}{7776}\langle \overline{q}q\rangle ^{2}\delta _{ab}  \notag \\
&& -\frac{x^{4}\langle
\overline{q}q\rangle \langle g_{s}^{2}G^{2}\rangle }{27648}\delta
_{ab}+\cdots.  \notag \\
&&  \label{eq:qProp}
\end{eqnarray}%
For the heavy quark $Q$, we use the propagator $S_{Q}^{ab}(x)$ as
\begin{eqnarray}
&&S_{Q}^{ab}(x)=i\int \frac{d^{4}k}{(2\pi )^{4}}e^{-ikx}\Bigg \{\frac{\delta
_{ab}\left( {\slashed k}+m_{Q}\right) }{k^{2}-m_{Q}^{2}}  \notag \\
&&+\frac{\langle g_{s}^{2}G^{2}\rangle}{12}\delta _{ab}m_{Q}\frac{k^{2}+m_{Q}{\slashed k}}{%
(k^{2}-m_{Q}^{2})^{4}}+\frac{\langle g_{s}^{3}G^{3}\rangle}{48}\delta _{ab}\frac{\left( {%
\slashed k}+m_{Q}\right) }{(k^{2}-m_{Q}^{2})^{6}}  \notag \\
&&\times \left[ {\slashed k}\left( k^{2}-3m_{Q}^{2}\right) +2m_{Q}\left(
2k^{2}-m_{Q}^{2}\right) \right] \left( {\slashed k}+m_{Q}\right) +\cdots %
\Bigg \}.  \notag \\
&&  \label{eq:QProp}
\end{eqnarray}
Beside the above expressions, which contain the perturbative or free light and heavy propagators and various vacuum condensates, we have extra terms containing one gluon field strength tensor, as well. These terms are  $ -\frac{ig_{s}G_{ab}^{\mu \nu }}{32\pi ^{2}x^{2}}\left[ \slashed x\sigma
_{\mu \nu }+\sigma _{\mu \nu }\slashed x\right] $ and $ -\frac{g_{s}G_{ab}^{\alpha \beta }}{4}\frac{\sigma _{\alpha \beta }\left( {
\slashed k}+m_{Q}\right) +\left( {\slashed k}+m_{Q}\right) \sigma _{\alpha\beta }}{(k^{2}-m_{Q}^{2})^{2}}   $ to be replaced  in Eqs.\ (\ref{eq:OPE1}) -(\ref{eq:OPE4})  instead of  the light and heavy quark propagators,  respectively.  These terms, when multiplied to each other in the presence of vacuum, make another two-gluon condensates, which we take into account their contributions and they will be discussed later. 
Here, we use  the short-hand notations
\begin{eqnarray}
G_{ab}^{\alpha \beta } &=&G_{A}^{\alpha \beta
}\lambda _{ab}^{A}/2,\,\,~~G^{2}=G_{\alpha \beta }^{A}G_{\alpha \beta }^{A},  \notag
\\
G^{3} &=&\,\,f^{ABC}G_{\mu \nu }^{A}G_{\nu \delta }^{B}G_{\delta \mu }^{C},
\end{eqnarray}%
where $f^{ABC}$ are the structure constants of the color group $SU_{c}(3)$.

We will treat  $ \langle 0 |G^n_{\alpha \beta}(x)G^m_{\alpha' \beta'}(0)|0 \rangle $ in Eqs.\ (\ref{eq:OPE1}) -(\ref{eq:OPE4}) in two different forms: first we  displace it by the full propagator of the gluon in coordinate space,
\begin{eqnarray}
\label{eq:Gprop}
&&\langle 0 |G^n_{\alpha \beta}(x)G^m_{\alpha' \beta'}(0)||0 \rangle =
\frac{\delta^{mn}}{2 \pi^2 x^4} [g_{\beta \beta'}(g_{\alpha \alpha'}-\frac{4 x_{\alpha} x_{\alpha'}}{x^2}) \nonumber \\
&& +(\beta, \beta') \leftrightarrow (\alpha, \alpha')
 -\beta \leftrightarrow \alpha -\beta' \leftrightarrow \alpha'], \notag \\
 &&{}
\end{eqnarray}
and perform all the computations.  This is  equivalent to the diagrams at which the   valence-gluon  is a full propagator.  At the next step, we consider $ \langle 0 |G^n_{\alpha \beta}(x)G^m_{\alpha' \beta'}(0)|0 \rangle $ as the gluon condensate  and take  the first term of the Taylor expansion at $x=0$ , i.e.
\begin{eqnarray}
\label{eq:Gcond}
&&\langle 0 |G^n_{\alpha \beta}(0)G^m_{\alpha' \beta'}(0)|0 \rangle =
\frac{\langle G^2\rangle }{96}\delta^{mn} [g_{\alpha \alpha'} g_{\beta \beta'} \notag \\
&&-g_{\alpha \beta'} g_{\alpha'\beta }],
\end{eqnarray}
which  stands for the diagrams with the gluon interacting with the QCD vacuum.

By substituting Eq.\ (\ref{eq:Gprop}) or (\ref{eq:Gcond}) in Eqs.\ (\ref{eq:OPE1}) -(\ref{eq:OPE4}), we need to use
\begin{eqnarray}
\label{eq:tntn}
&&\mathrm t^{n}[a,b] \mathrm t^{n}[a^{\prime },b^{\prime }]=\frac{1}{2}\left(\delta^{ab'}\delta^{a'b}-\frac{1}{3}\delta^{ab}\delta^{a'b'}\right).
\end{eqnarray}

The desired QCD sum rules for the physical observables  can be obtained using the same Lorentz structures in physical and OPE sides. In the case under consideration these structures are $\sim g_{\mu \nu }$ and $\sim I$. We choose the terms $\sim g_{\mu \nu }$ and $\sim I$ which represent contributions of V(AV) and S(PC) particles, respectively. The invariant amplitudes in $\Pi _{\mu \nu }^{\prime\mathrm{OPE}}(q)$ and $\Pi _{\mu \nu }^{\mathrm{OPE}}(q)$ corresponding to the structure $g_{\mu \nu }$ and $I$ in our following analysis will be denoted by $\Pi _{\mu \nu }^{\prime\mathrm{OPE}}(q^{2})$ and $\Pi _{\mu \nu }^{\mathrm{OPE}}(q^{2})$.To derive sum rules, we equate invariant amplitudes $\Pi ^{\mathrm{Phys}}(q^{2})$ and $\Pi ^{\mathrm{OPE}}(q^{2})$ corresponding to these structures, and apply the
Borel transformation to both sides of the obtained sum rules. The last
operation is necessary to suppress contributions stemming from the higher
resonances and continuum states. At the following phase of manipulations, we make use of an
assumption about the quark-hadron duality as well, and subtract from the physical
side contributions of higher resonances and continuum terms. After these manipulations, the final sum rule equality acquires a
dependence on the Borel $M^{2}$ and continuum threshold  $s_{0}$
parameters for each case. This equality, and second expression obtained by applying the
operator $d/d(-1/M^{2})$ to its both sides, form a system which is solved  to
find sum rules for the mass $m_{H}$ and current coupling $f_{H}$
\begin{equation}
m_{H}^{2}=\frac{d/d(-1/M^{2})[\Pi (M^{2},s_{0})]}{\Pi (M^{2},s_{0})},  \label{eq:Mass}
\end{equation}%
\begin{equation}
f_{H}^{2}=\frac{e^{m_{H}^{2}/M^{2}}}{m_{H}^{2}}\Pi (M^{2},s_{0}),  \label{eq:Coupling}
\end{equation}%
for each case.

In Eqs.\ (\ref{eq:Mass}) and (\ref{eq:Coupling}) the function $\Pi
(M^{2},s_{0})$  is the Borel transformed and continuum subtracted invariant
amplitude $\Pi ^{\mathrm{OPE}}(q^{2})$ for each channel. In the present work, we calculate $\Pi (M^{2},s_{0})$ by taking into account quark, gluon and mixed vacuum condensates up to
dimension $10$. It has the following form
\begin{equation}
\Pi (M^{2},s_{0})=\int_{m_{Q}^{2}}^{s_{0}}ds\rho ^{\mathrm{OPE}%
}(s)e^{-s/M^{2}}+\Pi (M^{2}),  \label{eq:InvAmp}
\end{equation}%
where $\rho ^{\mathrm{OPE}}(s)$ is the two-point spectral density. The
second component of the invariant amplitude $\Pi (M^{2})$ contains
nonperturbative contributions calculated directly from  the OPE side of each case. The explicit expressions of the functions $ \rho ^{\mathrm{OPE}}(s) $ and $\Pi (M^{2})$ are shown in the  Appendix,  as examples, for the  charm-nonstrange hybrid meson with the quantum numbers $J^{P(C)}=1^{-(-)}$.

\section{Numerical Results}
\label{sec:Numeric}
Now, we proceed to numerically analyze the obtained sum rules to get the values of the physical quantities for different channels. The quark, gluon and mixed condensates which enter to the sum rules (\ref%
{eq:Mass}) and (\ref{eq:Coupling}) are universal parameters of the computations:
\begin{eqnarray}
&&\langle \overline{q}q\rangle =-(0.24\pm 0.01)^{3}~\mathrm{GeV}^{3},\
\notag \\
&&\langle \overline{s}s\rangle =0.8~\langle \overline{q}q\rangle ,\
\notag \\
&&\langle \overline{q}g_{s}\sigma Gq\rangle =m_{0}^{2}\langle \overline{q}%
q\rangle ,\ m_{0}^{2}=(0.8\pm 0.1)~\mathrm{GeV}^{2},\   \notag \\
&&\langle \frac{\alpha _{s}G^{2}}{\pi }\rangle =(0.012\pm 0.004)~\mathrm{GeV}%
^{4},  \notag \\
&&\langle g_{s}^{3}G^{3}\rangle =(0.57\pm 0.29)~\mathrm{GeV}^{6},  \notag \\
&&m_{c}=1.275\pm 0.025~\mathrm{GeV}, \notag \\
&&m_{b}=4.18^{+0.03}_{-0.02}~\mathrm{GeV},  \notag \\
&&m_{s}=93^{+11}_{-5}~\mathrm{MeV}.  \label{eq:Parameters}
\end{eqnarray}%

The vacuum condensates have fixed numerical values, whereas the Borel and
continuum threshold parameters $M^{2}$ and $s_{0}$ can be varied within some regions, which have
to satisfy the standard constraints of the sum rules computations. The continuum threshold at each channel depends on the energy of the first excited state and it  is chosen like that the integrals do not receive contributions from the excited states when the ground state is studied. Unfortunately, we do not have experimental  information on the spectrum of the hybrid mesons under study. Hence, we apply some conditions to fix their working windows. The main restrictions that simultaneously  are applied are reaching to maximum possible pole contribution ($\mathrm{PC}$), convergence of the OPE and relatively weak dependence of the results on auxiliary parameters.  The $\mathrm{PC}$ is defined as 
\begin{equation}
\mathrm{PC}=\frac{\Pi(M^{2},\ s_{0})}{\Pi(M^{2},\ \infty )}.  \label{eq:Cond1}
\end{equation}
For the  OPE, we impose the condition of the higher the dimension of the perturbative and non-perturbative operators entering the calculations, the lower their contributions.  For the higher dimensional terms, i.e., $\mathrm{DimN=Dim(8+9+10)}$ we demand
\begin{equation}
R(M_{\mathrm{min}}^{2})=\frac{\Pi^{\mathrm{DimN}}(M_{\mathrm{%
min}}^{2},\ s_{0})}{\Pi(M_{\mathrm{min}}^{2},\ s_{0})}<0.01,
\label{eq:Cond2}
\end{equation}
which helps us to fix $M_{\mathrm{min}}^{2}$.  As we mentioned above, the last condition is the relatively weak dependence of the physical quantities to the helping parameters $M^{2}$ and $s_{0}$.  By applying all  these constraints we fix the working windows of the $M^{2}$ and $s_{0}$ that their full intervals will be shown in the tables containing the values of the  mass and current coupling  for all the channels under consideration. But before that, let us present the values of   $\mathrm{PC}$  for  different $J^{P(C)}$ states of heavy-light hybrid mesons at average value of $s_{0}$ and $[M_{\mathrm{min}}^{2},M_{\mathrm{max}}^{2}]$ in Tables\ \ref{C0_PCresults_table}-\ref{Bs_PCresults_table}. From these tables, we observe that the values of    $\mathrm{PC}$ at lower and higher limits of $ M^2 $ are obtained to be very high at bottom channels compared to the charmed ones. However, at the worst case which belongs to the charm-nonstrange hybrid mesons with   $1^{-(-)}$ the  $\mathrm{PC}$ is obtained to be $ 20\% $ at $ M_{\mathrm{max}}^{2} $, which is acceptable for the exotic  hadrons. 

\begin{widetext}

\begin{table}[!ht]
\centering
\caption{Values of PC for different quantum numbers of the ground state charm-nonstrange hybrid mesons. }
\label{C0_PCresults_table}
\begin{tabular}{cccc}
  $J^{P(C)}$ & $s_{0}(average)\ (GeV^2)$ & $[M_{\mathrm{min}}^{2},M_{\mathrm{max}}^{2}]\ (GeV^2)$
  & $\mathrm{PC}$ \\ 
\hline
  $0^{+(+)}$ & $26.0$ & $[6.0,7.0]$ & $ [0.36,0.22] $\\
  $0^{+(-)}$ & $22.0$ & $[5.0,7.0]$ & $ [0.64,0.37] $\\
  $0^{-(-)}$ & $32.0$ & $[6.0,8.0]$ & $ [0.50,0.35] $\\
  $0^{-(+)}$ & $22.0$ & $[5.0,7.0]$ & $ [0.66,0.39] $\\
  $1^{+(+)}$ & $22.0$ & $[5.0,7.0]$ & $ [0.41,0.21] $\\
  $1^{+(-)}$ & $22.0$ & $[5.0,7.0]$ & $ [0.55,0.30] $\\
  $1^{-(-)}$ & $22.0$ & $[5.0,7.0]$ & $ [0.38,0.20] $\\
  $1^{-(+)}$ & $22.0$ & $[5.0,7.0]$ & $ [0.55,0.30] $ 
\end{tabular}
\end{table}

\begin{table}[!ht]
\centering
\caption{Values of PC for different quantum numbers of the ground state charm-strange hybrid mesons. }
\label{Cs_PCresults_table}
\begin{tabular}{cccc}
  $J^{P(C)}$ & $s_{0}(average)\ (GeV^2)$ & $[M_{\mathrm{min}}^{2},M_{\mathrm{max}}^{2}]\ (GeV^2)$
  & $\mathrm{PC}$ \\ 
\hline
  $0^{+(+)}$ & $28.0$ & $[6.0,8.0]$ & $ [0.46,0.29] $\\
  $0^{+(-)}$ & $24.0$ & $[5.0,7.0]$ & $ [0.69,0.42] $\\
  $0^{-(-)}$ & $34.0$ & $[6.0,8.0]$ & $ [0.55,0.40] $\\
  $0^{-(+)}$ & $24.0$ & $[5.0,7.0]$ & $ [0.72,0.45] $\\
  $1^{+(+)}$ & $24.0$ & $[5.0,7.0]$ & $ [0.49,0.28] $\\
  $1^{+(-)}$ & $24.0$ & $[5.0,7.0]$ & $ [0.62,0.35] $\\
  $1^{-(-)}$ & $24.0$ & $[5.0,7.0]$ & $ [0.47,0.26] $\\
  $1^{-(+)}$ & $24.0$ & $[5.0,7.0]$ & $ [0.63,0.36] $
\end{tabular}
\end{table}

\begin{table}[!ht]
\centering
\caption{Values of PC for different quantum numbers of the ground state bottom-nonstrange hybrid mesons. }
\label{B0_PCresults_table}
\begin{tabular}{cccc}
   $J^{P(C)}$ & $s_{0}(average)\ (GeV^2)$ & $[M_{\mathrm{min}}^{2},M_{\mathrm{max}}^{2}]\ (GeV^2)$
  & $\mathrm{PC}$ \\ 
\hline
  $0^{+(+)}$ & $100.0$ & $[10.0,14.0]$ & $ [0.91,0.72] $\\
  $0^{+(-)}$ & $95.0$ & $[10.0,14.0]$ & $ [0.91,0.70] $\\
  $0^{-(-)}$ & $75.0$ & $[10.0,14.0]$ & $ [0.53,0.33] $\\
  $0^{-(+)}$ & $95.0$ & $[10.0,14.0]$ & $ [0.92,0.72] $\\
  $1^{+(+)}$ & $95.0$ & $[10.0,14.0]$ & $ [0.88,0.65] $\\
  $1^{+(-)}$ & $95.0$ & $[10.0,14.0]$ & $ [0.89,0.67] $\\
  $1^{-(-)}$ & $95.0$ & $[10.0,14.0]$ & $ [0.87,0.64] $\\
  $1^{-(+)}$ & $95.0$ & $[10.0,14.0]$ & $ [0.89,0.67] $
\end{tabular}
\end{table}

\begin{table}
\centering
\caption{Values of PC for different quantum numbers of the ground state bottom-strange hybrid mesons. }
\label{Bs_PCresults_table}
\begin{tabular}{cccccc}
 $J^{P(C)}$ & $s_{0}(average)\ (GeV^2)$ & $[M_{\mathrm{min}}^{2},M_{\mathrm{max}}^{2}]\ (GeV^2)$
  & $\mathrm{PC}$ \\ 
\hline
  $0^{+(+)}$ & $105.0$ & $[10.0,14.0]$ & $ [0.94,0.77] $\\
  $0^{+(-)}$ & $100.0$ & $[10.0,14.0]$ & $ [0.93,0.74] $\\
  $0^{-(-)}$ & $80.0$ & $[10.0,14.0]$ & $ [0.61,0.40] $\\
  $0^{-(+)}$ & $100.0$ & $[10.0,14.0]$ & $ [0.94,0.77] $\\
  $1^{+(+)}$ & $100.0$ & $[10.0,14.0]$ & $ [0.91,0.71] $\\
  $1^{+(-)}$ & $100.0$ & $[10.0,14.0]$ & $ [0.92,0.72] $\\
  $1^{-(-)}$ & $100.0$ & $[10.0,14.0]$ & $ [0.90,0.70] $\\
  $1^{-(+)}$ & $100.0$ & $[10.0,14.0]$ & $ [0.92,0.73] $
\end{tabular}
\end{table}

\end{widetext}

The mass $m_{H}$ and current coupling $f_{H}$ are demonstrated as functions of the Borel and continuum threshold parameters in Figs.\ \ref{fig:Mass} and  \ref{fig:Decay constant}. We observe that the results depict good stability against the changes in the value of the  auxiliary  parameter $M^2$ in its working interval. The residual dependence of the mass and current coupling on  $s_0$  constitutes the main source of the uncertainties that remain below the limits allowed by the method. 

\begin{widetext}

\begin{figure}[h!]
\begin{center}
\includegraphics[totalheight=6cm,width=8cm]{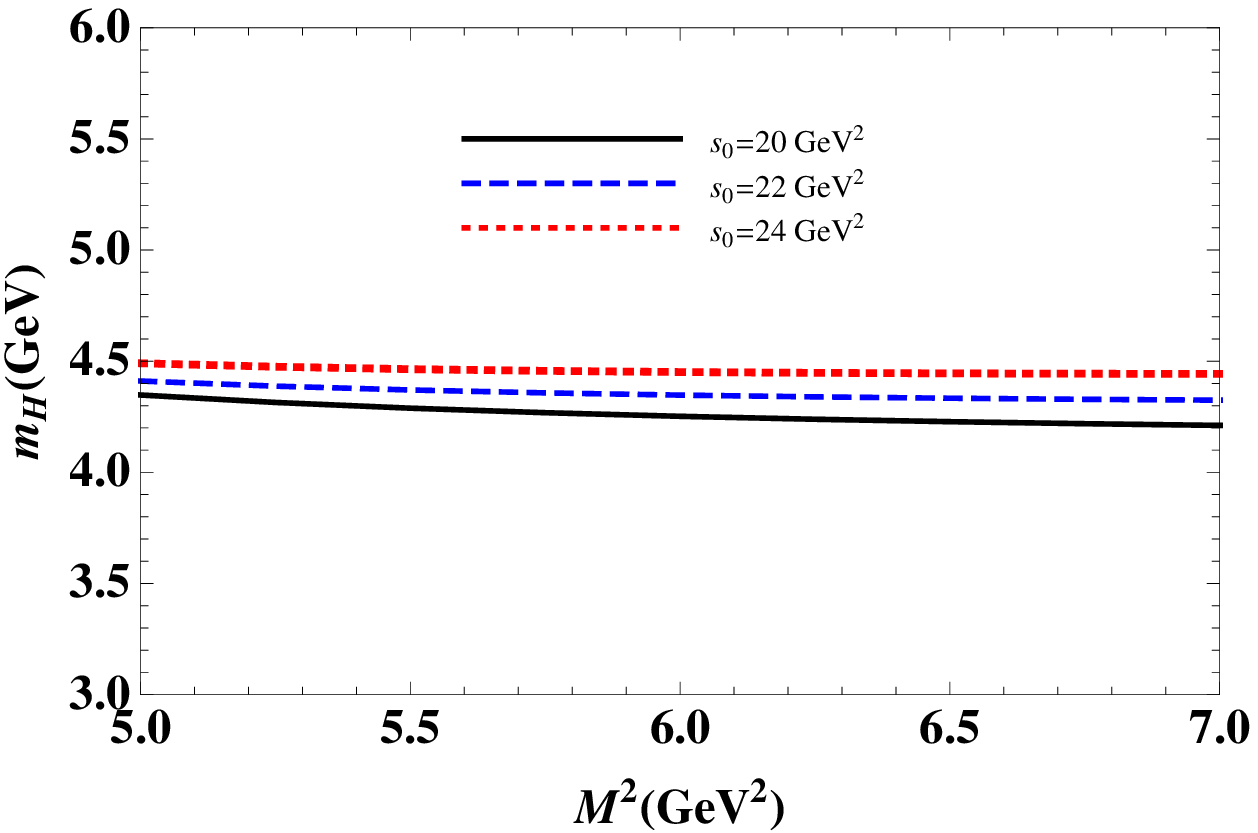}\,\, %
\includegraphics[totalheight=6cm,width=8cm]{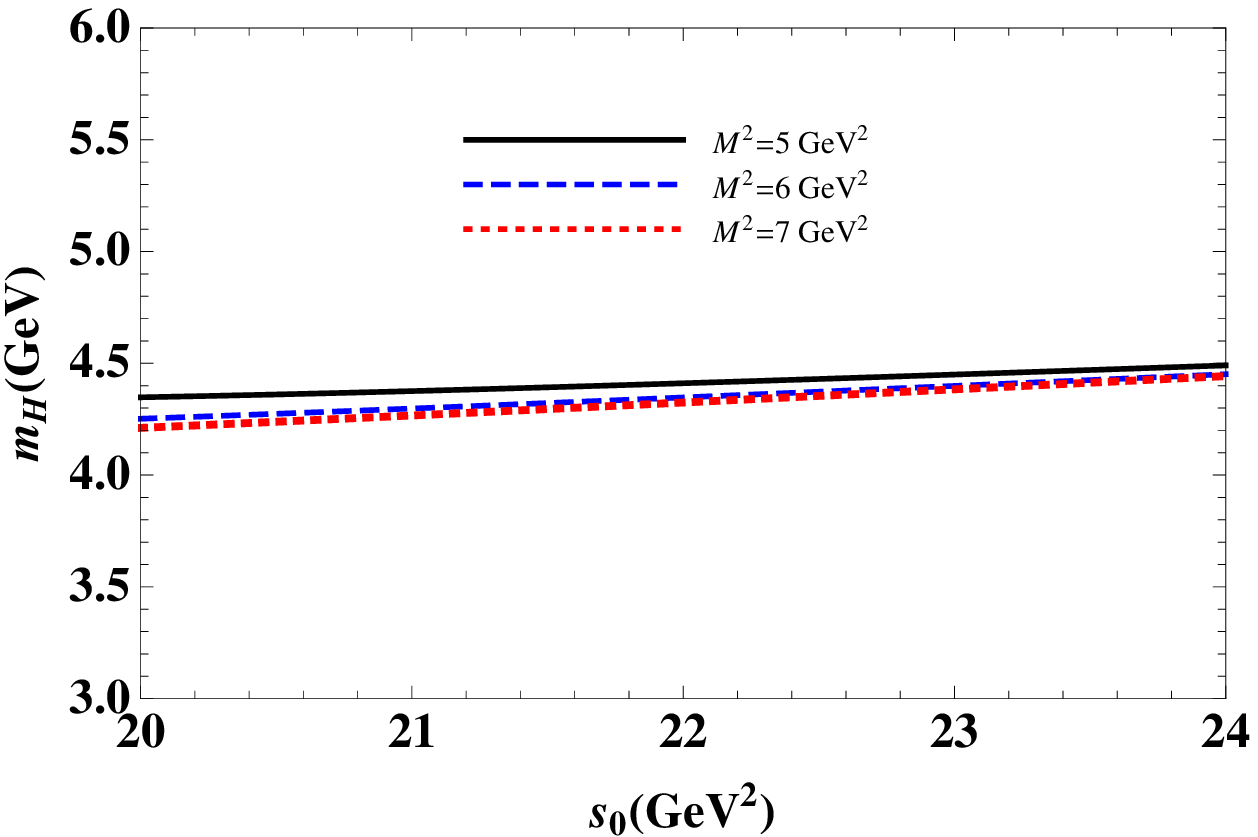}
\end{center}
\caption{ The mass of the charm-nonstrange hybrid meson with the quantum numbers $J^{P(C)}=1^{-(-)}$ as a function of the Borel parameter $M^2$ at fixed values of $s_0$ (left panel), and as a function
of the continuum threshold $s_0$ at fixed values of  $M^2$ (right panel).}
\label{fig:Mass}
\end{figure}

\begin{figure}[h!]
\begin{center}
\includegraphics[totalheight=6cm,width=8cm]{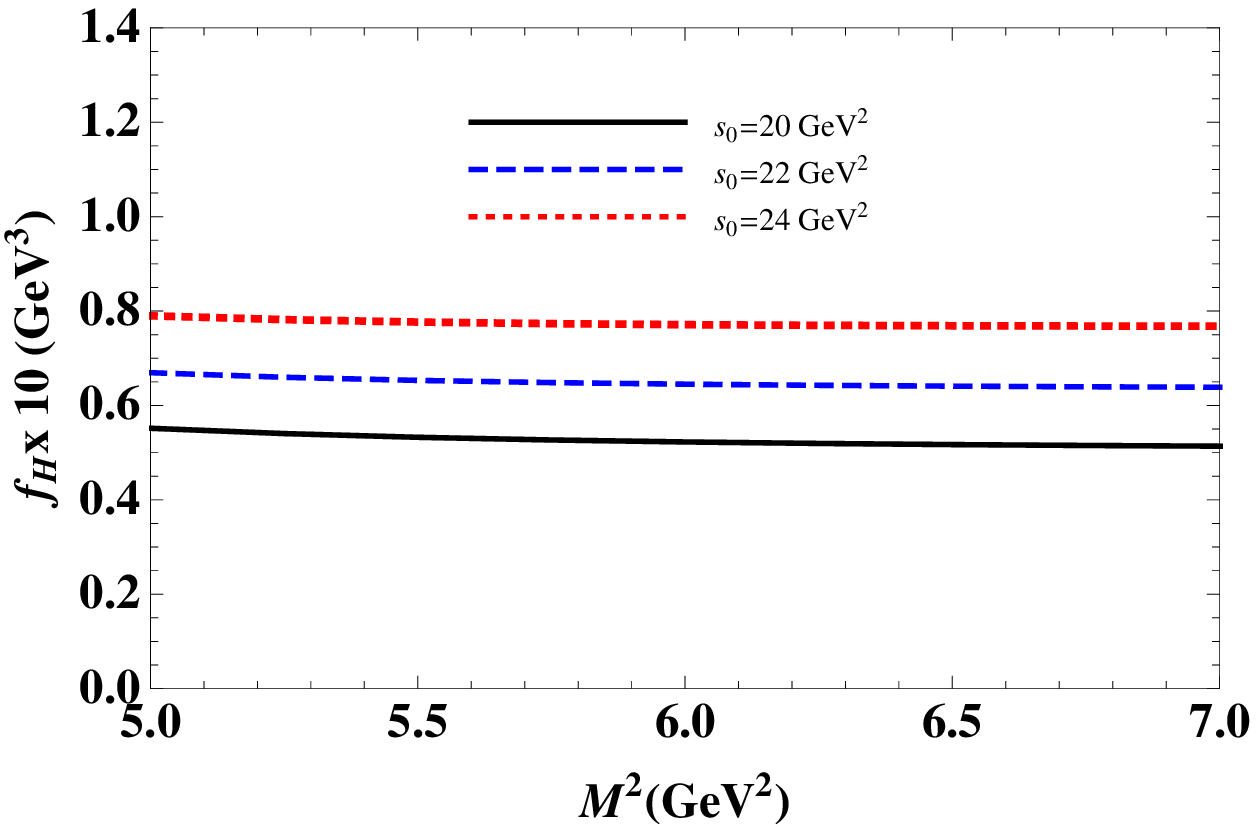}\,\, %
\includegraphics[totalheight=6cm,width=8cm]{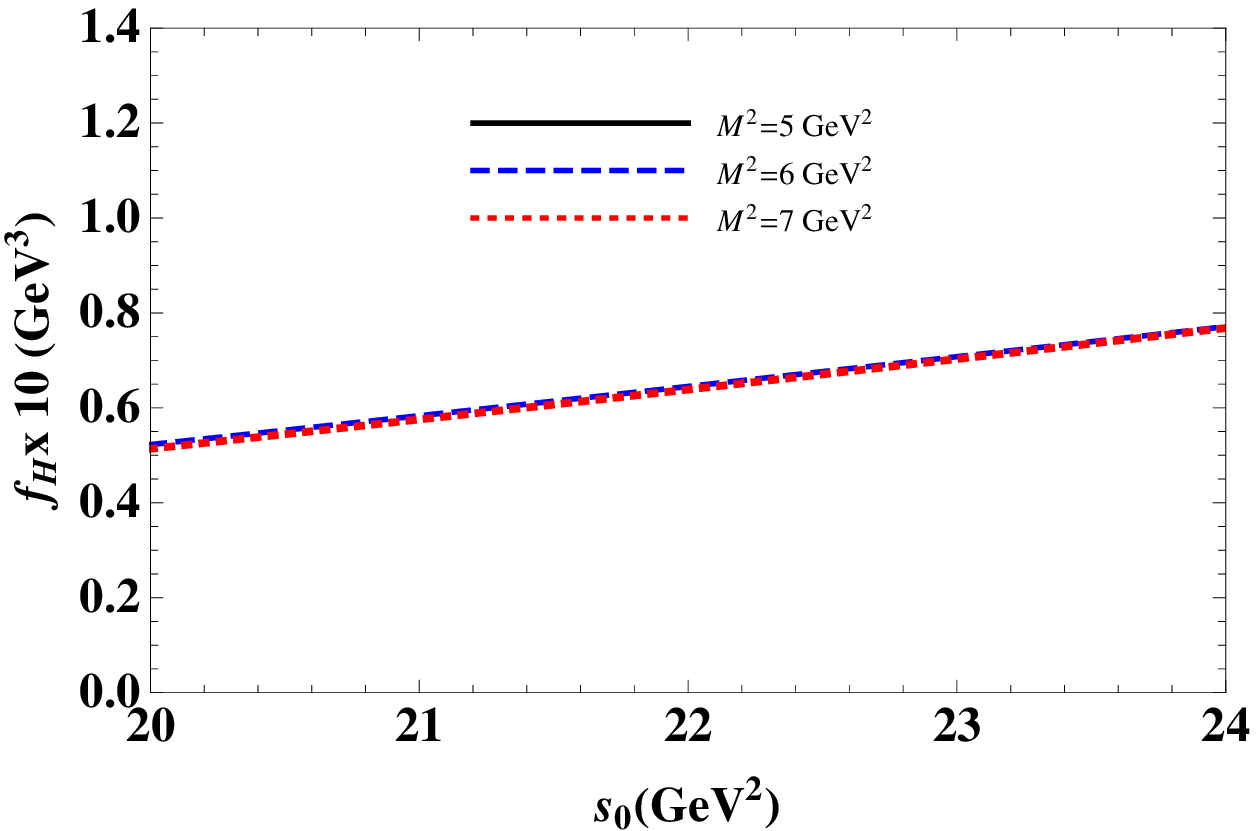}
\end{center}
\caption{ The current coupling of the charm-nonstrange hybrid meson with the quantum numbers $J^{P(C)}=1^{-(-)}$ as a function of the
Borel parameter $M^2$ at fixed values of $s_0$ (left panel), and as a function
of the continuum threshold $s_0$ at fixed values of  $M^2$ (right panel).}
\label{fig:Decay constant}
\end{figure}

\end{widetext}

Obtained from our analyses, the QCD sum-rules results for the mass and current coupling of  the heavy-light hybrid mesons are depicted in Tables\ \ref{C0_results_table}-\ref{Bs_results_table}.  The presented uncertainties are related to the errors in the calculations of the working regions for the auxiliary parameters as well as those exist in the values of other input parameters. Comparing our results with those of the existing predictions in  Ref. \cite{Ho:2016owu}, made by Laplace sum rules,  i.e., states with $J^{P(C)}\in \{0^{+(+)},1^{-(-)},1^{+(-)}\}$,  our results on the masses are well consistent with the predictions of Ref. \cite{Ho:2016owu}. However, our predictions for the current couplings differ considerably with the predictions of  Ref. \cite{Ho:2016owu} using Laplace sum rules for the mentioned quantum numbers. Let us compare the predictions of two studies for example for the $ 1^{-(-)} $ quantum numbers and different quark contents in  Table\ \ref{comparison}.  As is seen from this table, the values of the masses are in nice consistencies  between the present work and  Ref. \cite{Ho:2016owu} within the presented errors. In the case of current couplings, the results of  Ref. \cite{Ho:2016owu}  were presented dimensionless and  without uncertainties, but when we change them to $ GeV^3 $, the results of  Ref. \cite{Ho:2016owu} are about (2-4) times greater than our predictions.  As we also  previously mentioned, Ref. \cite{Ho:2016owu} finds unstable and non-reliable results for the states with $J^{P(C)}\in \{0^{+(-)},1^{+(+)},1^{-(+)}\}$.     The current couplings are among main input parameters to investigate the interactions of heavy-light hybrid mesons with other particles as well as their strong, electromagnetic and weak decays. 

\begin{widetext}

\begin{table}[!ht]
\centering
\caption{Values of the mass and current coupling for different quantum numbers of the ground state charm-nonstrange hybrid mesons. }
\label{C0_results_table}
\begin{tabular}{ccccc}
  $J^{P(C)}$ & $M^2 \pm \delta M^2\ (GeV^2)$
  & $s_0 \pm \delta s_0\ (GeV^2)$  & $m_H \pm \delta m_{H}\ (GeV)$ & $(f_{H}\pm \delta f_{H})\times10\ (GeV^3)$\\ 
\hline
  $0^{+(+)}$ & $7.0\pm 1.0$ & $26.0\pm 2.0$  & $ 4.86\pm 0.15$ & $ 0.68\pm 0.13$\\
  $0^{+(-)}$ & $6.0\pm 1.0$ & $22.0\pm 2.0$  & $ 3.73\pm 0.25$ & $ 0.62\pm 0.10$\\
  $0^{-(-)}$ & $7.0\pm 1.0$ & $32.0\pm 2.0$  & $ 5.50\pm 0.26$ & $ 1.13\pm 0.28$\\
  $0^{-(+)}$ & $6.0\pm 1.0$ & $22.0\pm 2.0$  & $ 3.73\pm 0.23$ & $ 0.66\pm 0.10$\\
  $1^{+(+)}$ & $6.0\pm 1.0$ & $22.0\pm 2.0$  & $ 4.22\pm 0.15$ & $ 0.64\pm 0.13$\\
  $1^{+(-)}$ & $6.0\pm 1.0$ & $22.0\pm 2.0$  & $ 3.88\pm 0.23$ & $ 0.73\pm 0.14$\\
  $1^{-(-)}$ & $6.0\pm 1.0$ & $22.0\pm 2.0$  & $ 4.36\pm 0.15$ & $ 0.65\pm 0.14$\\
  $1^{-(+)}$ & $6.0\pm 1.0$ & $22.0\pm 2.0$  & $ 3.91\pm 0.22$ & $ 0.74\pm 0.13$
\end{tabular}
\end{table}

\begin{table}[!ht]
\centering
\caption{Values of the mass and current coupling for different quantum numbers of the ground state charm-strange hybrid mesons. }
\label{Cs_results_table}
\begin{tabular}{ccccc}
    $J^{P(C)}$ & $M^2 \pm \delta M^2\ (GeV^2)$
  & $s_0 \pm \delta s_0\ (GeV^2)$  & $m_H \pm \delta m_{H}\ (GeV)$ & $(f_{H}\pm \delta f_{H})\times10\ (GeV^3)$\\ 
\hline
  $0^{+(+)}$ & $7.0\pm 1.0$ & $28.0\pm 2.0$  & $ 4.91\pm 0.14$ & $ 0.82\pm 0.13$\\
  $0^{+(-)}$ & $6.0\pm 1.0$ & $24.0\pm 2.0$  & $ 3.88\pm 0.25$ & $ 0.67\pm 0.11$\\
  $0^{-(-)}$ & $7.0\pm 1.0$ & $34.0\pm 2.0$  & $ 5.61\pm 0.25$ & $ 1.23\pm 0.31$\\
  $0^{-(+)}$ & $6.0\pm 1.0$ & $24.0\pm 2.0$  & $ 3.85\pm 0.24$ & $ 0.76\pm 0.11$\\
  $1^{+(+)}$ & $6.0\pm 1.0$ & $24.0\pm 2.0$  & $ 4.38\pm 0.13$ & $ 0.77\pm 0.14$\\
  $1^{+(-)}$ & $6.0\pm 1.0$ & $24.0\pm 2.0$  & $ 4.05\pm 0.23$ & $ 0.83\pm 0.15$\\
  $1^{-(-)}$ & $6.0\pm 1.0$ & $24.0\pm 2.0$  & $ 4.44\pm 0.13$ & $ 0.76\pm 0.13$\\
  $1^{-(+)}$ & $6.0\pm 1.0$ & $24.0\pm 2.0$  & $ 4.03\pm 0.22$ & $ 0.86\pm 0.15$
\end{tabular}
\end{table}

\begin{table}[!ht]
\centering
\caption{Values of the mass and current coupling for different quantum numbers of the ground state bottom-nonstrange hybrid mesons. }
\label{B0_results_table}
\begin{tabular}{ccccc}
    $J^{P(C)}$ & $M^2 \pm \delta M^2\ (GeV^2)$
  & $s_0 \pm \delta s_0\ (GeV^2)$  & $m_H \pm \delta m_{H}\ (GeV)$ & $(f_{H}\pm \delta f_{H})\times10\ (GeV^3)$\\  
\hline
  $0^{+(+)}$ & $12.0\pm 2.0$ & $100.0\pm 5.0$  & $ 8.44\pm 0.25$ & $ 2.48\pm 0.45$\\
  $0^{+(-)}$ & $12.0\pm 2.0$ & $95.0\pm 5.0$  & $ 7.92\pm 0.38$ & $ 2.07\pm 0.51$\\
  $0^{-(-)}$ & $12.0\pm 2.0$ & $75.0\pm 5.0$  & $ 8.31\pm 0.28$ & $ 1.58\pm 0.45$\\
  $0^{-(+)}$ & $12.0\pm 2.0$ & $95.0\pm 5.0$  & $ 7.84\pm 0.38$ & $ 2.10\pm 0.51$\\
  $1^{+(+)}$ & $12.0\pm 2.0$ & $95.0\pm 5.0$  & $ 8.24\pm 0.30$ & $ 2.65\pm 0.57$\\
  $1^{+(-)}$ & $12.0\pm 2.0$ & $95.0\pm 5.0$  & $ 8.08\pm 0.35$ & $ 2.56\pm 0.61$\\
  $1^{-(-)}$ & $12.0\pm 2.0$ & $95.0\pm 5.0$  & $ 8.32\pm 0.26$ & $ 2.70\pm 0.54$\\
  $1^{-(+)}$ & $12.0\pm 2.0$ & $95.0\pm 5.0$  & $ 8.07\pm 0.34$ & $ 2.58\pm 0.35$
\end{tabular}
\end{table}

\begin{table}
\centering
\caption{Values of the mass and current coupling for different quantum numbers of the ground state bottom-strange hybrid mesons. }
\label{Bs_results_table}
\begin{tabular}{ccccc}
    $J^{P(C)}$ & $M^2 \pm \delta M^2\ (GeV^2)$
  & $s_0 \pm \delta s_0\ (GeV^2)$  & $m_H \pm \delta m_{H}\ (GeV)$ & $(f_{H}\pm \delta f_{H})\times10\ (GeV^3)$\\ 
\hline
  $0^{+(+)}$ & $12.0\pm 2.0$ & $105.0\pm 5.0$  & $ 8.50\pm 0.26$ & $ 2.74\pm 0.51$\\
  $0^{+(-)}$ & $12.0\pm 2.0$ & $100.0\pm 5.0$  & $ 8.04\pm 0.39$ & $ 2.18\pm 0.55$\\
  $0^{-(-)}$ & $12.0\pm 2.0$ & $80.0\pm 5.0$  & $ 8.46\pm 0.27$ & $ 1.80\pm 0.48$\\
  $0^{-(+)}$ & $12.0\pm 2.0$ & $100.0\pm 5.0$  & $ 7.92\pm 0.39$ & $ 2.31\pm 0.57$\\
  $1^{+(+)}$ & $12.0\pm 2.0$ & $100.0\pm 5.0$  & $ 8.35\pm 0.30$ & $ 2.92\pm 0.62$\\
  $1^{+(-)}$ & $12.0\pm 2.0$ & $100.0\pm 5.0$  & $ 8.20\pm 0.35$ & $ 2.78\pm 0.67$\\
  $1^{-(-)}$ & $12.0\pm 2.0$ & $100.0\pm 5.0$  & $ 8.41\pm 0.28$ & $ 2.92\pm 0.59$\\
  $1^{-(+)}$ & $12.0\pm 2.0$ & $100.0\pm 5.0$  & $ 8.16\pm 0.36$ & $ 2.81\pm 0.67$
\end{tabular}
\end{table}

\begin{table}
\centering
\caption{Comparison of the values for the mass and current coupling for quantum numbers $J^{P(C)}=1^{-(-)}$ of the different heavy-light hybrid mesons between the present work (PW) and Ref.  \cite{Ho:2016owu}. The masses are  in $ GeV $ and the current couplings are in $ GeV^3 $.}
\label{comparison}
\begin{tabular}{ccccc}
    $J^{P(C)}=1^{-(-)}$ & $m_H  $ (PW)
  & $f_{H} \times10$ (PW) & $m_H  $ \cite{Ho:2016owu}  & $f_{H}\times10$  \cite{Ho:2016owu}\\ 
\hline
  charm-nonstrange & $ 4.36\pm 0.15$ & $ 0.65\pm 0.14$  & $ 4.40\pm 0.19$  & $ 2.99$\\
  charm-strange & $ 4.44\pm 0.13$ & $ 0.76\pm 0.13$  & $ 4.28\pm 0.19$  & $ 2.60$\\
 bottom-nonstrange & $ 8.32\pm 0.26$ & $ 2.70\pm 0.54$  & $ 8.74\pm 0.25$  & $ 8.87$\\
  bottom-strange & $ 8.41\pm 0.28$ & $ 2.92\pm 0.59$  & $ 8.46\pm 0.32$  & $ 6.72$
\end{tabular}
\end{table}
\begin{table}
\centering
\caption{Comparison of the values for the mass and current coupling for states with quantum numbers $J^{P(C)}=1^{-(-)}$ of the different heavy-light hybrid mesons obtained by including non-perturbative effects up to six and ten dimensions. The  primed masses and decay constants belong to the results up to dimension six and those without prime show the results up to ten mass dimensions. The masses are  in $ GeV $ and the current couplings are in $ GeV^3 $.}
\label{comparison2}
\begin{tabular}{ccccc}
    $J^{P(C)}=1^{-(-)}$ & $m_{H'}  $ (PW)
  & $f_{H'} \times10$ (PW) & $m_H  $ (PW)
  & $f_{H} \times10$ (PW) \\ 
\hline
  charm-nonstrange & $ 4.32\pm 0.14$ & $ 0.64\pm 0.13$ & $ 4.36\pm 0.15$ & $ 0.65\pm 0.14$  \\
  charm-strange & $ 4.40\pm 0.13$ & $ 0.75\pm 0.13$ & $ 4.44\pm 0.13$ & $ 0.76\pm 0.13$  \\
 bottom-nonstrange & $ 8.30\pm 0.28$ & $ 2.69\pm 0.55$ & $ 8.32\pm 0.26$ & $ 2.70\pm 0.54$ \\
  bottom-strange & $ 8.39\pm 0.29$ & $ 2.90\pm 0.61$ & $ 8.41\pm 0.28$ & $ 2.92\pm 0.59$ 
\end{tabular}
\end{table}
\begin{figure}[h!]
\begin{center}
\includegraphics[totalheight=6cm,width=8cm]{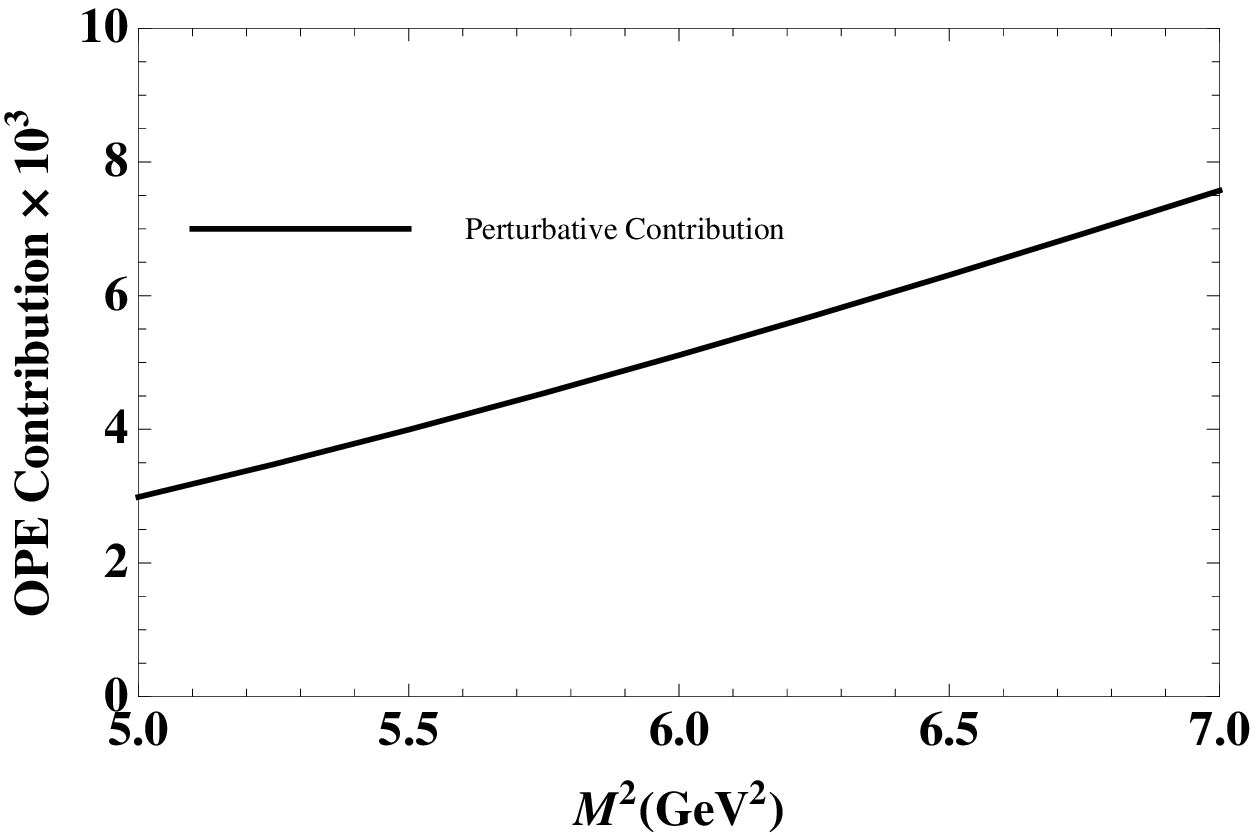}\,\, %
\includegraphics[totalheight=6cm,width=8cm]{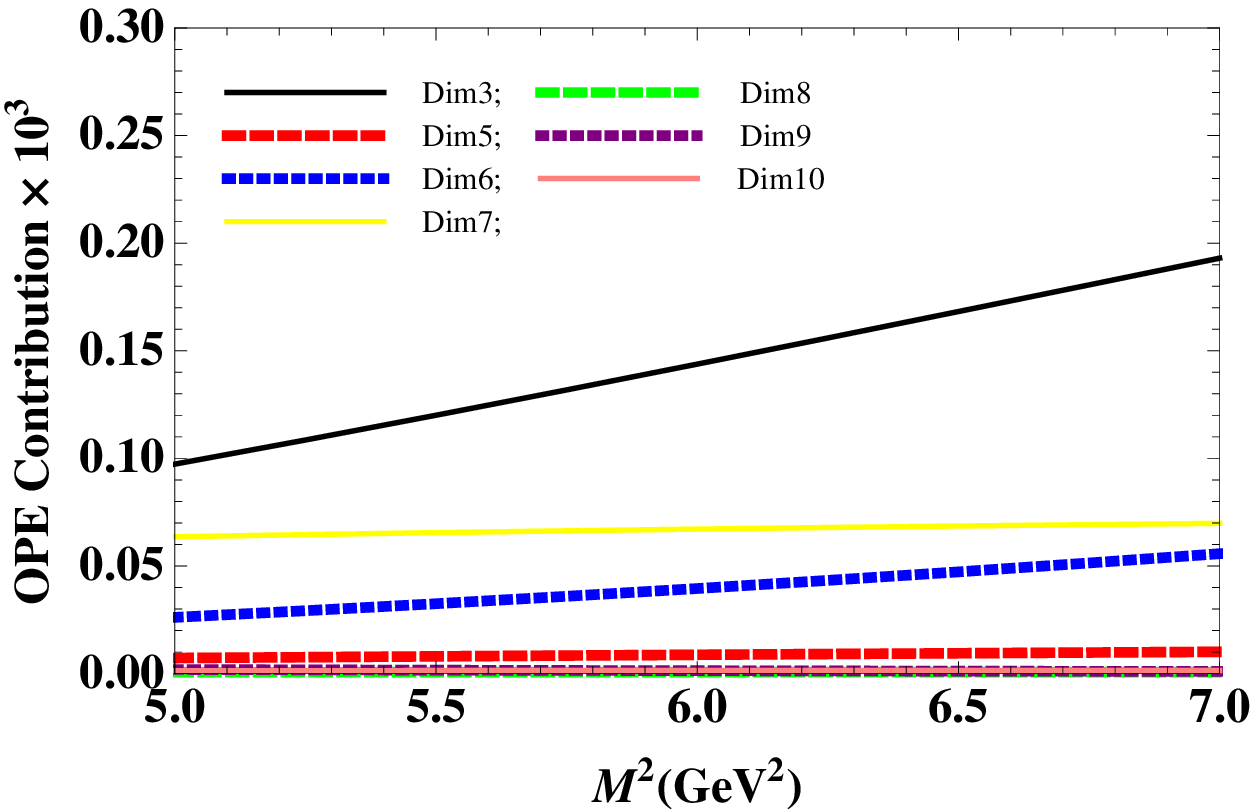}
\end{center}
\caption{The perturbative and non-perturbative contributions  in units of  $ GeV^8 $ to the sum rule of the charm-nonstrange hybrid meson with the quantum numbers $J^{P(C)}=1^{-(-)}$: Perturbative contribution  (left panel) and absolute values of different non-perturbative contributions (right panel)  as functions of $M^2$   at  average $s_0$. }
\label{fig:Contribution}
\end{figure}

\end{widetext}

Regarding the comparison of our results with those of Ref. \cite{Ho:2016owu} and our reason for making improvements on the subject, we would like to present more details. As we previously mentioned, Ref. \cite{Ho:2016owu}  consider the analyses up to six-dimension non-perturbative operators, while we make our analyses considering these effects up to ten mass dimensions.  Despite, our results on the masses of some states are in nice consistency with those of Ref. \cite{Ho:2016owu} , the values of decay constants obtained in these two studies are very different and Ref. \cite{Ho:2016owu}  finds the results of masses and decay constants for some states to be unstable. Now, a question is raised: Do these conclusions are still valid when we consider the analyses up to six dimensional non-perturbative operators? What are the contributions of higher dimensional operators? To answer this questions let us first compare our results for instance for $J^{P(C)}=1^{-(-)}$  states  obtained by including non-perturbative effects up to six and ten dimensions in  Table \ref{comparison2}.  As is seen from this table,  the higher dimensions (seven to ten non-pertubative) show very small contributions and the results with six and ten dimensions do not differ considerably. As we previously mentioned, we have a nice convergence of the OPE and it is also clear from  figure \ref{fig:Contribution}, which depicts the  perturbative and different non-perturbative contributions for the sum rules of the charm-nonstrange hybrid meson with the quantum numbers $J^{P(C)}=1^{-(-)}$, as an example. This figure shows that the main contribution comes from the perturbative part and the higher dimensional non-perturbative operators have very small contributions referring to the nice convergence of the OPE in the analyses. Considering these evidences,  the differences between our results and those of Ref. \cite{Ho:2016owu}  can be attributed to different methods of  analyses that these two studies use.

\begin{widetext}

\end{widetext}

\section{Summary and Conclusion}
\label{sec:Conc}
In this article, we  investigated the  the mass and current coupling of the scalar, pseodoscalar, vector and axial-vector heavy-light hybrid mesons with different quantum numbers and quark contents  employing the QCD two-point Borel sum rule method by choosing appropriate interpolating currents. We restricted the auxiliary parameters entered the sum rules based on the standard prescriptions of the method and included into analyses the quark-gluon condensations up to dimension 10. We got stable and reliable results for the mass and current couplings for all the considered quantum numbers and quark contents with respect to the variations of the auxiliary parameters in their working windows. We extracted the values of the mass and current coupling for various  $J^{P(C)}$ and quark contents and presented them in  Tables\ \ref{C0_results_table}-\ref{Bs_results_table}. Our predictions for the mass of the states with $J^{P(C)}\in \{0^{+(+)},1^{-(-)},1^{+(-)}\}$  are in good agreements with the existing predictions of the Ref. \cite{Ho:2016owu} made using Laplace sum rules. The obtained results for the current couplings of the hybrid mesons with $J^{P(C)}\in \{0^{+(+)},1^{-(-)},1^{+(-)}\}$  differ considerably with those of the  Ref. \cite{Ho:2016owu} when they are written in the same dimensions.  Comparison of the theoretical results on various parameters of the hybrid states obtained in the present study  with possible  future experimental data will shed light on the nature, quark-gluon organization and quantum numbers of these states. The obtained values for the spectroscopic parameters of the heavy-light hybrid mesons with different spin-parities can be used in investigation of their strong, electromagnetic and weak decays as well as their various interactions with other particles. 

\section*{ACKNOWLEDGMENTS}
K. Azizi is thankful to Iran Science Elites Federation (Saramadan)
for the partial  financial support provided under the grant number ISEF/M/400150.

\appendix*

\begin{widetext}

\section{ Invariant amplitude $\Pi
(M^{2},s_{0})$ for the  charm-nonstrange hybrid meson with the quantum numbers $J^{P(C)}=1^{-(-)}$}

\renewcommand{\theequation}{\Alph{section}.\arabic{equation}} \label{sec:App}
The invariant amplitude $\Pi (M^{2},s_{0})$ obtained after the Borel
transformation and subtraction procedures is given by Eq.\ (\ref{eq:InvAmp}). 
Here, we present the explicit expressions of the spectral density $\rho ^{\mathrm{OPE}}(s)$ and the function $\Pi
(M^{2})$ as an example  for the  charm-nonstrange hybrid meson with the quantum numbers $J^{P(C)}=1^{-(-)}$, which are written as 
\begin{equation}
\rho ^{\mathrm{OPE}}(s)=\rho ^{\mathrm{pert.}}(s)+\sum_{N=3}^{6}\rho ^{%
\mathrm{DimN}}(s),\ \ \Pi (M^{2})=\sum_{N=6}^{10}\Pi ^{\mathrm{DimN}}(M^{2}),
\label{eq:A1}
\end{equation}%
respectively. The components of $\rho ^{\mathrm{OPE}}(s)$ and $\Pi (M^{2})$
are given by the expressions%
\begin{equation}
\rho ^{\mathrm{DimN}}(s)=\int_{0}^{1}d%
\alpha \rho ^{\mathrm{DimN}}(s,\alpha ),  \label{eq:A2}
\end{equation}%
and
\begin{equation}
\Pi ^{\mathrm{DimN}%
}(M^{2})=\int_{0}^{1}d\alpha \Pi ^{\mathrm{DimN}}(M^{2},\alpha ).
\label{eq:A4}
\end{equation}%
depending on whether $\rho $ and $\Pi (M^{2})$ are functions of $\alpha $. In Eqs. (\ref{eq:A2}) and (\ref{eq:A4})
the variable $\alpha $ is the Feynman parameter.

The perturbative and nonperturbative components of the spectral density $%
\rho ^{\mathrm{pert.}}(s,\alpha )$ and $\rho ^{\mathrm{Dim3(4,5,6)%
}}(s,\alpha )$ \ have the forms:
\begin{eqnarray}
&&\rho ^{\mathrm{pert.}}(s,\alpha )=\frac{g_{s}^{2}(m_{Q}^{2}+sL)^{2}\alpha ^{2}[6m_{q}m_{Q}(-3+2\alpha)+\alpha[m_{Q}^{2}(-10+3\alpha)+2s(14-17\alpha +3\alpha ^{2})]]}{2^{9}\cdot 3^{2}\pi
^{4}L^{2}} ,
\end{eqnarray}%
\begin{equation}
\rho ^{\mathrm{Dim3}}(s,\alpha )=\frac{g_{s}^{2}\langle \overline{q}q\rangle\alpha[-2m_{Q}(m_{Q}^{2}+sL)+m_{Q}[m_{Q}^{2}(-5+3\alpha)+4s(2-3\alpha+\alpha ^{2})]]}{2^{5}\cdot 3\pi
^{2}} ,
\end{equation}%
\begin{eqnarray}
&&\rho ^{\mathrm{Dim4}}(s,\alpha )=-\frac{\langle \alpha _{s}G^{2}/\pi
\rangle g_{s}^{2}\alpha ^{2}[-16m_{q}m_{Q}L\alpha +sL^{2}(3+8\alpha)+m_{Q}^{2}(-3+37\alpha -6\alpha ^{2})]}{2^{11}\cdot 3^{3}\pi ^{2}L^{2}}\notag \\
&&+\frac{1}{2^{5}\cdot 3}\langle \alpha _{s}G^{2}/\pi \rangle [3m_{q}m_{Q}+(3m_{Q}^{2}+4sL)\alpha] ,
\end{eqnarray}%
\begin{eqnarray}
&&\rho ^{\mathrm{Dim5}}(s,\alpha )=\frac{g_{s}^{2}m_{o}^{2}\langle \overline{q}q\rangle[3m_{Q}L+m_{q}(-5+11\alpha -6\alpha ^{2})]}{2^{6}\cdot 3^{2}\pi ^{2}} ,
\end{eqnarray}%
\begin{eqnarray}
&&\rho ^{\mathrm{Dim6}}(s,\alpha )=\frac{g_{s}^{4}\langle \overline{q}q\rangle ^{2}(5-11\alpha +6\alpha ^{2})}{2^{4}\cdot 3^{5}\pi ^{2}} ,
\end{eqnarray}%

Components of the function $\Pi (M^{2})$ are:%
\begin{eqnarray}
&&\Pi ^{\mathrm{Dim6}}(M^{2},\alpha )=\frac{g_{s}^{2}\langle g_{s}^{3}G^{3} \rangle \alpha ^{2}m_{Q}^{2}}{2^{5}\cdot 3L^{5}\pi ^{4}} \exp %
\left[ \frac{m_{Q}^{2}}{M^{2}L}\right] 
\Bigg \{-\frac{m_{q}m_{Q}^{3}}{2^{3}\cdot 3\cdot 5M^{4}L}-\frac{m_{q}m_{Q}^{3}}{2^{4}\cdot 3\cdot 5M^{4}}-\frac{m_{q}m_{Q}}{2^{2}\cdot 3\cdot 5M^{2}}-\frac{m_{Q}^{4}\alpha}{2^{5}\cdot 3\cdot 5M^{4}L^{2}}  \notag \\
&&+\frac{m_{q}m_{Q}^{3}\alpha }{3^{2}\cdot 5M^{4}L}-\frac{m_{Q}^{4}\alpha}{2^{5}\cdot 5M^{4}L}-\frac{m_{Q}^{2}\alpha}{2^{5}\cdot 3\cdot 5M^{2}L}+\frac{\alpha}{2^{5}\cdot 3\cdot 5}+\frac{m_{q}m_{Q}^{3}\alpha }{2^{3}\cdot 3^{2} \cdot 5M^{4}}+\frac{7m_{q}m_{Q}\alpha }{2^{2}\cdot 3^{2} \cdot 5M^{2}}-\frac{m_{Q}^{2}\alpha }{2^{5}\cdot 3 \cdot 5M^{2}}\notag \\
&&+\frac{19m_{Q}^{4}\alpha ^{2} }{2^{6}\cdot 3^{2} \cdot 5M^{4}L^{2}}-\frac{7m_{q}m_{Q}^{3}\alpha ^{2} }{2^{3}\cdot 3^{2} \cdot 5M^{4}L}+\frac{7m_{Q}^{4}\alpha ^{2} }{2^{6}\cdot 3 \cdot 5M^{4}L}+\frac{17m_{Q}^{2}\alpha ^{2} }{2^{6}\cdot 3^{2} \cdot 5M^{2}L}-\frac{\alpha ^{2} }{2^{5}\cdot 5}-\frac{m_{q}m_{Q}\alpha ^{2} }{2^{2}\cdot 3^{2}M^{2}}+\frac{m_{Q}^{2}\alpha ^{2} }{2^{6}\cdot 3\cdot 5M^{2}}\notag \\
&&-\frac{7m_{Q}^{4}\alpha ^{3}}{2^{6}\cdot 3 \cdot 5M^{4}L^{2}}+\frac{m_{q}m_{Q}^{3}\alpha ^{3}}{2^{2}\cdot 3^{2} \cdot 5M^{4}L}-\frac{m_{Q}^{4}\alpha ^{3}}{2^{6}\cdot 3 \cdot 5M^{4}L}-\frac{m_{Q}^{2}\alpha ^{3}}{2^{6}\cdot 3M^{2}L}+\frac{\alpha ^{3}}{2^{5}\cdot 5}+\frac{m_{q}m_{Q}\alpha ^{3} }{2^{2}\cdot 3^{2}\cdot 5M^{2}}+\frac{m_{Q}^{2}\alpha ^{3} }{2^{6}\cdot 3\cdot 5M^{2}}\notag \\
&&+\frac{m_{Q}^{4}\alpha ^{4} }{2^{6}\cdot 5M^{4}L^{2}}+\frac{m_{Q}^{2}\alpha ^{4} }{2^{6}\cdot 3\cdot 5M^{2}L}-\frac{\alpha ^{4} }{2^{5}\cdot 3\cdot 5}-\frac{m_{Q}^{4}\alpha ^{5} }{2^{6}\cdot 3^{2}\cdot 5M^{4}L^{2}}+\frac{m_{Q}^{2}\alpha ^{5} }{2^{6}\cdot 3^{2}\cdot 5M^{2}L} \Bigg \} ,
\end{eqnarray}%

\begin{eqnarray}
&&\Pi ^{\mathrm{Dim7}}(M^{2},\alpha )=\frac{\langle \alpha _{s}G^{2}/\pi
\rangle \langle \overline{q}q \rangle}{2^{3}\cdot 3} \exp %
\left[ -\frac{m_{Q}^{2}}{M^{2}}\right]\left\{-\frac{g_{s}^{2}m_{Q}}{2^{3}\cdot 3^{2}\pi}+\pi ^{2}(\frac{m_{q}}{3}+m_{Q}+\frac{m_{q}m_{Q}^{2}}{2\cdot 3M^{2}})\right\} ,
\end{eqnarray}%

\begin{eqnarray}
&&\Pi ^{\mathrm{Dim8}}(M^{2},\alpha )=\frac{\langle \alpha
_{s}G^{2}/\pi \rangle ^{2} \pi ^{2}}{2^{4}\cdot 3^{2}L} \exp %
\left[ \frac{m_{Q}^{2}}{M^{2}L}\right] 
\Bigg \{\frac{m_{q}m_{Q}^{3}}{2^{2}M^{4}L^{3}}+\frac{m_{q}m_{Q}}{2M^{2}L^{2}}
+\frac{5m_{Q}^{2}}{2^{5}\cdot 3M^{2}L}-\frac{1}{2^{3}\cdot 3}  \notag \\
&&-\frac{m_{q}m_{Q}^{3}\alpha }{2M^{4}L^{3}}-\frac{m_{Q}^{4}\alpha}{2^{2}\cdot 3M^{4}L^{3}} 
-\frac{m_{q}m_{Q}\alpha}{M^{2}L^{2}}+\frac{m_{Q}^{2}\alpha}{2^{2}\cdot 3M^{2}L^{2}}-\frac{5m_{Q}^{2}\alpha}{2^{5}\cdot 3M^{2}L}+\frac{\alpha}{2^{3}\cdot 3}+\frac{m_{q}m_{Q}^{3}\alpha ^{2}}{2^{2}M^{4}L^{3}}  \notag \\
&&+\frac{m_{Q}^{4}\alpha ^{2}}{2^{2}\cdot 3M^{4}L^{3}}+\frac{m_{q}m_{Q}\alpha ^{2}}{2M^{2}L^{2}}-\frac{m_{Q}^{2}\alpha ^{2}}{2^{2}\cdot 3M^{2}L^{2}} \Bigg \} ,
\end{eqnarray}%

\begin{eqnarray}
&&\Pi ^{\mathrm{Dim9}}(M^{2},\alpha )=\frac{\langle \alpha _{s}G^{2}/\pi
\rangle \langle \overline{q}q \rangle m_{o}^{2}}{2^{3}\cdot 3M^{2}}\exp %
\left[ \frac{m_{Q}^{2}}{M^{2}L}\right] 
\Bigg \{\frac{g_{s}^{2}}{L^{2}}(-\frac{m_{q}m_{Q}^{4}}{2^{4}\cdot 3^{2}M^{4}L^{2}}-\frac{m_{Q}^{3}}{2^{4}\cdot 3^{2}M^{2}L^{2}}-\frac{m_{q}m_{Q}^{2}}{2^{2}\cdot 3^{3}M^{2}L}-\frac{m_{Q}}{2^{3}\cdot 3^{2}L}  
\notag \\
&&+\frac{5m_{q}m_{Q}^{4}\alpha }{2^{4}\cdot 3^{3}M^{4}L^{2}}+\frac{m_{Q}^{3}\alpha}{2^{2}\cdot 3^{2}M^{2}L^{2}} 
+\frac{m_{q}m_{Q}^{2}\alpha}{2^{2}\cdot 3^{3}M^{2}L}+\frac{m_{Q}\alpha}{2^{3}\cdot 3L}+\frac{m_{Q}\alpha}{2^{3}\cdot 3^{2}}-\frac{m_{q}m_{Q}^{4}\alpha ^{2}}{2^{3}\cdot 3^{3}M^{4}L^{2}}-\frac{5m_{Q}^{3}\alpha ^{2}}{2^{4}\cdot 3^{2}M^{2}L^{2}}  \notag \\
&&-\frac{m_{Q}\alpha ^{2}}{2^{3}\cdot 3L}+\frac{m_{Q}\alpha ^{2}}{2^{2}\cdot 3^{2}}+\frac{m_{Q}^{3}\alpha ^{3}}{2^{3}\cdot 3^{2}M^{2}L^{2}}+\frac{m_{Q}\alpha ^{3}}{2^{3}\cdot 3^{2}L})-\pi ^{2}(\frac{m_{q}m_{Q}^{4}}{2^{2}\cdot 3^{2}M^{4}}-\frac{m_{Q}}{M^{2}}+\frac{13m_{q}m_{Q}^{2}}{2^{2}\cdot 3^{2}M^{2}}-\frac{29m_{q}}{2^{2}\cdot 3^{2}}) \Bigg \} ,
\end{eqnarray}%

\begin{eqnarray}
&&\Pi ^{\mathrm{Dim10}}(M^{2},\alpha )=\frac{\langle \alpha _{s}G^{2}/\pi
\rangle}{2^{3}\cdot 3^{3}M^{2}}\exp %
\left[ \frac{m_{Q}^{2}}{M^{2}L}\right] 
\Bigg \{\dfrac{\langle \overline{q}q \rangle ^{2}g_{s}^{4}}{3^{4}L^{3}}(\frac{m_{Q}^{4}}{2^{2}M^{4}L}+\frac{m_{Q}^{2}}{3M^{2}}-\frac{5m_{Q}^{4}\alpha}{2^{2}\cdot 3M^{4}L}-\frac{m_{Q}^{2}\alpha}{3M^{2}}  
\notag \\
&&+\frac{m_{Q}^{4}\alpha ^{2}}{2\cdot 3M^{4}L})+\frac{\langle \overline{q}q \rangle ^{2}\pi ^{2}g_{s}^{2}}{3^{4}}(\frac{m_{Q}^{4}}{M^{4}}-\frac{13m_{Q}^{2}}{M^{2}}+29 )+\frac{g_{s}^{3}G^{3}}{M^{4}L^{5}}(\frac{m_{q}m_{Q}^{5}}{2^{3}\cdot 5M^{2}L}+\frac{m_{q}m_{Q}^{5}}{2^{4}\cdot 5M^{2}L}+\frac{m_{q}m_{Q}^{3}}{2\cdot 5}+\frac{m_{Q}^{6}\alpha}{2^{5}\cdot 3\cdot 5M^{2}L^{2}}\notag \\
&&-\frac{m_{q}m_{Q}^{5}\alpha}{2^{2}\cdot 5M^{2}L}+\frac{m_{Q}^{6}\alpha}{2^{5}\cdot 5M^{2}L}+\frac{m_{Q}^{4}\alpha}{2^{3}\cdot 3\cdot 5L}-\frac{m_{q}m_{Q}^{3}\alpha}{5}+\frac{m_{Q}^{4}\alpha}{2^{5}\cdot 5}-\frac{m_{Q}^{6}\alpha ^{2}}{2^{5}\cdot 5M^{2}L^{2}}+\frac{m_{q}m_{Q}^{5}\alpha ^{2}}{2^{3}\cdot 5M^{2}L}-\frac{m_{Q}^{6}\alpha ^{2}}{2^{5}\cdot 5M^{2}L}-\frac{m_{Q}^{4}\alpha ^{2}}{2^{3}\cdot 5L}\notag \\
&&+\frac{m_{q}m_{Q}^{3}\alpha ^{2}}{2\cdot 5}-\frac{m_{Q}^{4}\alpha ^{2}}{2^{5}\cdot 5}+\frac{m_{Q}^{6}\alpha ^{3}}{2^{5}\cdot 5M^{2}L^{2}}+\frac{m_{Q}^{4}\alpha ^{3}}{2^{3}\cdot 5L}-\frac{m_{Q}^{6}\alpha ^{4}}{2^{5}\cdot 3\cdot 5M^{2}L^{2}}-\frac{m_{Q}^{4}\alpha ^{4}}{2^{3}\cdot 3\cdot 5L}) \Bigg \} ,
\end{eqnarray}%
where have used  the following short-hand notation:
\begin{eqnarray}
L=\alpha -1.\
\end{eqnarray}

\end{widetext}

\renewcommand{\theequation}{\Alph{section}.\arabic{equation}} \label{sec:App}



\end{document}